\documentclass[12pt,paper]{JHEP}          
\usepackage{epsfig}
\newcommand{\NPB}[3]{\emph{ Nucl.~Phys.} \textbf{B#1} (#2) #3}   
\newcommand{\PLB}[3]{\emph{ Phys.~Lett.} \textbf{B#1} (#2) #3}   
\newcommand{\PRD}[3]{\emph{ Phys.~Rev.} \textbf{D#1} (#2) #3}

\def\sla{\raise.15ex\hbox{$/$}\kern-.57em}

\let\a=\alpha  \let\b=\beta   \let\g=\gamma   \let\d=\delta   \let\e=\epsilon
       \let\q=\theta   \let\i=\iota    
\let\l=\lambda \let\m=\mu     \let\n=\nu      \let\x=\xi      \let\p=\pi     
\let\r=\rho    \let\s=\sigma  \let\t=\tau      \let\f=\phi   
    \let\y=\psi          
  \let\D=\Delta   \let\L=\Lambda
            
\let\F=\Phi                
\newcommand{\be}{\begin{equation}}
\newcommand{\ee}{\end{equation}}
\newcommand{\beba}{\begin{equation}\begin{array}{lcl}}
\newcommand{\eaee}{\end{array}\end{equation}}
\newcommand{\bea}{\begin{eqnarray}}
\newcommand{\eea}{\end{eqnarray}}
\newcommand{\ba}{\begin{array}}
\newcommand{\ea}{\end{array}}

\usepackage{psfig}
\usepackage{graphics}
\usepackage{feynmf}
\title{The anomalous magnetic moment of the muon in the 
D-brane realization of the Standard Model}
\author{E. Kiritsis$^{1,2}$, P. Anastasopoulos$^{1}$
\\
$^1$ Department of Physics, University of Crete and FORTH\\
P.O.Box 2208, GR-710 03 Heraklion, GREECE\\
\smallskip
$^2$ Laboratoire de Physique Th{\'e}orique Ecole Polytechnique\footnote{Unit{\'e}
mixte du CNRS and de l'Ecole Polytechnique, UMR 7644} \\
91128, Palaiseau, FRANCE\\
Email: {\tt kiritsis, panasta@physics.uoc.gr}}
\preprint{\hepph{0201295}\\
          S 002.0102}

\abstract{The anomalous magnetic moment of the muon is evaluated in the D-brane
realization of the Standard Model. It is pointed out that the massive anomalous
U(1) gauge bosons predicted, 
give extra contributions that are compatible with current experimental
data.}

\begin{document}

\maketitle
\section{Introduction}

The recent precise measurement of the anomalous magnetic moment (AMM) of
muon $\a_{muon}=(g-2)/2$ from the Brookhaven AGS experiment \cite{BNL} gave 
\be
\a_{muon}^{exp}=116 592 023 (151) \times 10^{-11}
\label{AMMexp}
\ee
The difference between the experimental value (\ref{AMMexp}) and the theoretical 
expectation, (for a review see \cite{theory}), due to standard model (SM) is\footnote{Recently there  has been
a reappraisal of the theoretical value \cite{corr} due to  a potential error
in the hadronic contribution. Taking this into account the  discrepancy
becomes $(25\pm 16) \times 10^{-10}$ namely 1.6 $\sigma$ away from the
experimental result.}   
\be
\d\a_{muon} = \a_{muon}^{exp}-\a_{muon}^{SM} = (43\pm 16)\times 10^{-10}.
\label{difference}
\ee
The experimental precision is unprecented and it is going to reach $\pm
4\times 10^{-10}$ soon.
It becomes thus important to examine the signals of physics  beyond the SM. 
Various explanations for a discrepancy have 
been proposed building on earlier computations \cite{early}. 
Many of those assume SUSY broken at a mass scale not far above the 
weak scale \cite{kane,feng,martinwells,Byrne,others}. 
Other approaches include large or warped extra dimensional models,
extended gauge structure and other alternatives \cite{Nath,Graesser,Park,Song}. 

In this paper, we are pointing out that the experimental result (\ref{AMMexp}) appears 
naturally in the D-brane realization of the SM in the context of orientifold vacua of 
string theory.
String theory vacua with a very low string scale $M_s$ \cite{ant,Ant,aadd,twoc,st}, and 
supersymmetry broken at that scale do not suffer directly from the ordinary hierarchy 
problem of the scalar masses \cite{add,aadd}. Rather, the hierarchy problem transmutes 
into the question as to why four-dimensional gravity is so weak. 
Moreover, if the string scale is around a few TeV, observation of novel effects at the 
near future experiments becomes a realistic possibility.
A low string scale compatible with the known value of the Planck scale can be easily 
accommodated in ground states of unoriented open and closed strings. Solvable vacua 
of this type are orientifolds of closed strings. Such vacua include various type of 
D-branes stretching their worldvolumes in the four non-compact dimensions while 
wrapping additional worldvolume dimensions around cycles of the compact six torus. 
Moreover, they include non-dynamical orientifold planes that cancel the charges of the 
D-branes, implementing the (un)orientability condition and stabilizing the vacuum 
(cancellation of tadpoles).
 
Gauge interactions are described by open strings whose ends are
confined on the D-branes, while gravity is mediated by closed strings in the bulk
\cite{aadd}. Ordinary matter is preferably generated by the fluctuations of the open strings 
and is thus also localized on the appropriate D-branes. The observed hierarchy between 
the Planck and the weak scale is accounted for by two or more large dimensions, transverse 
to our brane-world.
Since the masses of the SM matter are well below the string scale, 
the branes on which the SM fields 
are located must be close together in the internal transverse compact space.
Thus, for some questions, it makes sense to study this local configuration of 
branes without direct reference to the global groundstate configuration. 
Questions of ground state stability on the other hand are global questions.

In \cite{akt}, the local D-brane configuration that can reproduce the SM fields was presented.
The brane gauge-group is $U(3)\times U(2)\times U(1)$ and strong and electroweak apartinteractions 
arise from two different collections of coincident branes, leading to different gauge couplings.
The hypercharge is a linear combination of the abelian factors of the gauge group. All such 
hypercharge embeddings were classified in \cite {akt}. The abelian gauge symmetries orthogonal 
to the hypercharge are ``anomalous", their anomaly cancelled by a Green-Schwarz mechanism 
\cite{sagnotti,ibanez}. Such gauge symmetries are broken at the string scale, the gauge 
bosons becoming massive. The electroweak gauge symmetry is broken by the vacuum expectation 
values of two Higgs doublets, which are both necessary to give masses to all quarks and leptons. 
Moreover, the minimal spectrum  can be completely non-supersymmetric.
In such a case matching with the SM predicts that the string scale is in the range 6-8 TeV 
\cite{akt}.
Several orientifold models as well as local brane configurations that come close to the SM have 
been described in the literature \cite{dsm}.

In the gauge sector the only free parameters, not fixed by the anomalies, are the masses of the 
two anomalous gauge bosons. These are of the order of the string scale. Their precise values can 
be obtained by a one-loop string calculation \cite{akr} once a concrete string model exists.
Such masses are proportional to $g_i M_s$, where the $g_i$ are the gauge couplings, consequently  
we expect that they are in the region of a few TeV.

In this minimal realization, the only extra low-lying states from the standard model fields 
are the two massive ``anomalous" gauge bosons\footnote{The ``heavy" states
comprise oscillator states with masses of the order of the string scale.
There are also some KK excitations with comparable masses.}. 
The particles couple minimally to the leptons with 
strengths that are fixed by the known gauge couplings.
Thus, they provide computable contributions to the anomalous magnetic moment of the muon, the only 
uncertainty coming from the uncertainty in their masses\footnote{There is an extra uncertainty 
originating in the stringy corrections (contributions of higher oscillator modes of the open strings). 
We will discuss such uncertainties in the concluding section.} 

In this paper we compute such $(g-2)_{\rm anom}$ contributions and show that they are in the range 
implied by the experimental result. We use (\ref{AMMexp}) to provide precise constrains for the masses
of the anomalous U(1)'s in the TeV range.

The structure of this paper is as follows:
In section two we describe in more detail the main features of the D-brane realization of the SM.
In section three we describe the essential ingredients of the one-loop diagrammatic calculation of 
$(g-2)_{\rm anom}$. In section four the calculation of $(g-2)_{\rm anom}$ is done for the D-SM.
Section five contains our conclusions and further comments. 
In appendix A we describe the precise field theoretic 
Lagrangian of the D-SM. In appendix B details of the diagrammatic calculations and 
proof of the gauge invariance of the result can be found.

\section{The D-brane realization of SM}

The simplest way the $SU(3)\times SU(2)\times U(1)$ gauge group of the SM  can be embedded 
in a product of unitary 
groups appearing on D-brane stacks is as a subgroup of $U(3)\times U(2)\times U(1)
$\footnote{In fact the minimal 
embedding is in $U(3)\times U(2)$, however such an embedding has phenomenological problems:
 proton stability 
cannot be protected and some SM fields cannot get masses.}.
A U(n) factor arises from n coincident D-branes. As $U(3)=SU(3)\times U(1)$, 
a string with one end on this group 
of branes is a triplet under $SU(3)$ with $Q_3=\pm 1$ abelian charge. 
Thus, $Q_3$ is identified with the gauged 
baryon number. 
Similarly, the second factor arises from two coincident D-branes ("weak" branes) 
and the gauged overall abelian 
charge $Q_2$ is identified with the weak-doublet number. Both collections have 
their own gauge couplings $g_3$, 
$g_2$ that are functions of the string coupling $g_s$ and possible compactification 
volumes. 
The necessity for the extra $U(1)$ factor is due to the fact that we cannot express
 the hypercharge as a linear 
combination of baryon and weak-doublet numbers
\footnote{It turns out that a complete collection of SM D-branes 
(one that can accommodate all the endpoints of SM strings) includes a fourth U(1)$_b$ 
component that does not 
participate in the hypercharge. Such a D-brane wraps the large dimensions, and 
consequently its coupling is ultra 
weak. It is also anomalous and thus massive \cite{berlin}.
 Due to its weak coupling its contributions to magnetic 
moments are negligible compared to the ones we consider. We will thus ignore it in this paper.}.
The U(1) brane can be in principle independent of the other branes and  has in general a different 
gauge coupling $g_1$.
In \cite{akt}, the $U(1)$ brane has been put on top of either 
the color or the weak D-branes. Thus, $g_1$ is equal 
to either $g_3$ or $g_2$.

Let us denote by $Q_3$, $Q_2$ and $Q_1$ the three $U(1)$ charges 
of $U(3)\times U(2)\times U(1)$. These charges 
can be fixed so that they lead to the right hypercharge. In order that we can 
match the measured gauge couplings 
with the ones appropriate for the brane-configuration and also avoid hierarchy problems 
we find that we have to 
put the $U(1)$ brane on top of the color branes. Consequently we set $g_1=g_3$.
This fixes the string scale to be between 6 to 8 TeV \cite{akt}. 
There are two possibilities for charge assignments. Under 
$SU(3)\times SU(2)\times U(1)_3 \times U(1)_2 \times U(1)_1$
the members of a given quark and lepton family have the following quantum numbers:
\bea
&Q &({\bf 3},{\bf 2};1,1+2z,0)_{1/6}\nonumber\\
&u^c &({\bf\bar 3},{\bf 1};-1,0,0)_{-2/3}\nonumber\\
&d^c &({\bf\bar 3},{\bf 1};-1,0,1)_{1/3}\label{charges}\\
&L   &({\bf 1},{\bf 2};0,1,z)_{-1/2}\nonumber\\
&l^c &({\bf 1},{\bf 1};0,0,1)_1\nonumber\eea
where $z=0,-1$. From (\ref{charges}) and the requirement that the  Higgs doublet has
hypercharge 1/2, one finds two possible assignments for it:
\bea
H\ \ ({\bf 1},{\bf 2};0,1+2z,1)_{1/2}\quad &H'\ \ ({\bf 1},{\bf 2};0,-(1+2z),0)_{1/2}
\label{Higgs}
\eea  
The trilinear Yukawa terms are
\bea
\label{HY}
z=0  \ :\qquad H'Qu^c\ &,& \quad H^\dagger Ll^c
     \ ,\quad  H^\dagger Qd^c\\
z=-1 \ :\qquad H' Qu^c\ &,& \quad H'^\dagger Ll^c
     \ ,\quad  H^\dagger Qd^c
\label{HtildeY}
\eea
In each case, two Higgs doublets are necessary to give masses to all quarks
and leptons.
The U(3) and U(1) branes are D3 branes. The U(2) branes are D7 branes whose four
extra longitudinal directions are wrapped on a four-torus of volume 2.5 in
string units \cite{akt}. 
The spectator U(1)$_b$ brane is stretching in the bulk but the fermions that end on
it do not have KK excitations.
Thus, the only SM field that has KK excitations is a linear combination
of the hypercharge gauge boson and the two anomalous U(1) gauge bosons.
The masses of KK states, are shifted from the basic state by multiples of $0.8
M_s$.

We will now describe the structure of the gauge sector for the D-brane configuration above.
We denote by $A^i_\m$ the $U(1)_i$ gauge fields and $F^i_{\m\n}$ their corresponding field strengths. 
Also we denote $G^\b_{\m\n}$ the field strengths of the non-abelian gauge group where $\b$ 
runs over the two simple factors. There is also a set of two axion fields $b^\a$ with normalized
kinetic terms.
Starting from the kinetic terms of the gauge fields and requesting for the cancellation
of the $Q T^\a T^\a$ mixed anomalies, we can write down the most general low energy action
\bea
{\cal L} = &-&{1\over 4} \sum_i F^i_{\mu\nu}F^{i,\mu\nu}
             +\sum_i \bar{\y} Q_i \sla{A^i} \y  
             -{1\over 4} \sum_a Tr G^a_{\mu\nu}G^{a,\mu\nu}          \nonumber\\
           &+& \sum_{\a, \b} \L_{\a,\b}{b^\a \over M_s}\epsilon^{\mu\nu\rho\sigma}
               Tr[G^\b_{\mu\nu} G^{\b}_{\rho\sigma}]
             + \sum_\a (\partial_{\mu} b^\a-M_s\l^{\a i} A^i_{\mu})
                       (\partial^{\mu} b^\a-M_s\l^{\a j} A^{j,\mu})  \nonumber\\
           &+& \sum_{\a,i,j} {C_{\a ij}\over M_s} 
                     \epsilon^{\mu\nu\rho\sigma}\partial_{\mu} b^\a A^i_{\nu} 
                     F^j_{\rho\sigma}
             + \sum_{i,j,k} {D_{ijk}\over M_s}\epsilon^{\mu\nu\rho\sigma} 
           A^i_{\mu}A^j_{\nu} F^k_{\rho\sigma}                       \label{lor}\\
           &+& \sum_\a Z_\a{b^\a\over M_s}\epsilon^{\mu\nu\rho\sigma}
                  Tr[R_{\mu\nu} R_{\rho\sigma}]                      \nonumber\eea
where charge operators $Q_i$ contain all coupling dependence. 
The last term involves the curvature two-form $R_{\mu\nu}$ and 
is responsible for the cancellation of the gravitational anomalies.
Under $U(1)$ gauge transformations (modified by the anomaly) 
\bea
A^i_\m \rightarrow A^i_\m+\partial_\m \e^i \quad, \quad b^\a 
\rightarrow b^\a+\sum_i \l^{\a i} A^i_\m
\label{trns}\eea
we have
\bea
D_{ijk}=-D_{jik} \quad , \quad \sum_a \L_{\a,\b} \l^{\a,\i}=Tr[Q^iT_\b T_\b]
\eea
\bea
D_{ijk}=-\sum_a C_{\a ij} \l^{\a k}=Tr[Q^i Q^j Q^k] 
\quad, \quad \sum_a Z_\a \l^{\a, i}=Tr[Q^i]
\eea
The only free parameters which are not fixed by the anomalies are $\l^{\a i}$. These
define the mass matrix of gauge bosons $M^2_{ij}=M^2_s\l^{\a i}\l^{bj}$. This matrix 
is symmetric and has a zero eigenvalue corresponding to the non-anomalous hypercharge. 
The $\l^{\a i}$ can be computed by a string calculation.
The parameters remaining in the mass matrix is the $2\times 2$
submatrix of the anomalous gauge bosons.

Now, we will describe the couplings of the gauge fields in more details. 
The two first terms of (\ref{lor}) are written as
\be
{\cal L} = -{1\over 4} \sum_i F^iF^i +\sum_i {g_i \over \sqrt{i}} \bar{\y} Q_i \sla{A^i} \y
\ee
where $g_i$ are the $SU(i)$ coupling constants and the charges have the standard integral 
normalization (\ref{charges}). We will set
$x={g_3/\sqrt{3}\over g_2/\sqrt{2}}=\sqrt{5/3}$ as $g_2/g_3\sim \sqrt{0.4}$ \cite{akt}. 
Doing a $O(3)$ rotation, we can go to a basis where the kinetic 
terms of the $U(1)$ gauge fields are still diagonal, while one of them corresponds 
to the hypercharge: $A_i =U_{ij} \widetilde{A}_j $ with $A_Y = \widetilde{A}_1 $.
This rotation is different in each theory.\\

For the $z=0$ case we use
\be
U=
\left( 
\ba {ccc}
{2\sqrt{3}\over \sqrt{28+9x^2}} & 
-{\sqrt{16+9x^2}\sin{\q}\over \sqrt{28+9x^2}} &
{\sqrt{16+9x^2}\sin{\q}\sqrt{3}\over \sqrt{28+9x^2}} \\
-{3x\over \sqrt{28+9x^2}} & 
-{2(-2\sqrt{28+9x^2}\cos{\q} +3\sqrt{3}x\sin{\q})\over \sqrt{28+9x^2}\sqrt{16+9x^2}} & 
{2(2\sqrt{28+9x^2}\sin{\q} +3\sqrt{3}x\cos{\q})\over \sqrt{28+9x^2}\sqrt{16+9x^2}} \\
{4\over \sqrt{28+9x^2}} & 
{3x\sqrt{28+9x^2}\cos{\q} +8\sqrt{3}\sin{\q}\over \sqrt{28+9x^2}\sqrt{16+9x^2}} &
{3x\sqrt{28+9x^2}\sin{\q} -8\sqrt{3}\cos{\q}\over \sqrt{28+9x^2}\sqrt{16+9x^2}} 
\ea \right)
\label{Uarrey}
\ee
and the $U(1)$ charges:
\[
Q_{Y} \sim Q_1-{Q_2\over 2} + {2Q_3\over 3}
\]
\bea
Q_{\a}\sim -\sqrt{3}x(16+9x^2)\sin{\q}Q_1  
            +2(2\sqrt{28+9x^2}\cos{\q}-3\sqrt{3}x\sin{\q})Q_2+       \nonumber\\ 
     +(3x^2\sqrt{28+9x^2}\cos{\q}+8\sqrt{3}x\sin{\q})Q_3             \label{rotchargone} \eea
\bea        
Q_{b}\sim \sqrt{3}x(16+9x^2)\cos{\q}Q_1  
        +2(2\sqrt{28+9x^2}\sin{\q}+3\sqrt{3}x\cos{\q})Q_2+           \nonumber\\ 
        +(3x^2\sqrt{28+9x^2}\sin{\q}-8\sqrt{3}x\cos{\q})Q_3          \nonumber           \eea
We can obtain the $z=-1$ case from the one above by $x\rightarrow-x$. The matrix $U$ is now
\be
U=
\left( \ba {ccc}
{2\sqrt{3}\over \sqrt{28+9x^2}} & 
-{\sqrt{16+9x^2}\sin{\q}\over \sqrt{28+9x^2}} &
{\sqrt{16+9x^2}\sin{\q}\sqrt{3}\over \sqrt{28+9x^2}} \\
{3x\over \sqrt{28+9x^2}} & 
{2(2\sqrt{28+9x^2}\cos{\q} +3\sqrt{3}x\sin{\q})\over \sqrt{28+9x^2}\sqrt{16+9x^2}} & 
-{2(-2\sqrt{28+9x^2}\sin{\q} +3\sqrt{3}x\cos{\q})\over \sqrt{28+9x^2}\sqrt{16+9x^2}} \\
{4\over \sqrt{28+9x^2}} & 
{-3x\sqrt{28+9x^2}\cos{\q} +8\sqrt{3}\sin{\q}\over \sqrt{28+9x^2}\sqrt{16+9x^2}} &
-{3x\sqrt{28+9x^2}\sin{\q} +8\sqrt{3}\cos{\q}\over \sqrt{28+9x^2}\sqrt{16+9x^2}} \ea \right)
\label{Uarreysecond}
\ee
and the charges:
\[
Q_{Y} \sim Q_1+{Q_2\over 2} + {2Q_3\over 3}
\]
\bea
Q_{\a}\sim  -\sqrt{3}x(16+9x^2)\sin{\q}Q_1  
            +2(2\sqrt{28+9x^2}\cos{\q}+3\sqrt{3}x\sin{\q})Q_2+       \nonumber\\ 
            +(-3x^2\sqrt{28+9x^2}\cos{\q}+8\sqrt{3}x\sin{\q})Q_3     \label{rotchargtwo} \eea
\bea
Q_{b} \sim   \sqrt{3}x(16+9x^2)\cos{\q}Q_1  
            +2(2\sqrt{28+9x^2}\sin{\q}-3\sqrt{3}x\cos{\q})Q_2-       \nonumber\\ 
            -(3x^2\sqrt{28+9x^2}\sin{\q}-8\sqrt{3}x\cos{\q})Q_3      \nonumber           \eea
The parameter $\q$ can be used to diagonalize the mass matrix of the two anomalous $U(1)$s
$A_\a$ and $A_b$. The two eigenvalues $\m_\a^2$, $\m_b^2$ and $\q$ parametrize effectively 
the $2\times 2$ mass matrix. The masses of the anomalous $U(1)$ gauge fields have also 
contributions from the Higgs effect since the Higgses are also charged under the anomalous $U(1)$s. 
We evaluate these in appendix A. However, such corrections are of order of $m_Z/M_s$ and are thus 
subleading for our purposes.
String theory calculations indicate  that $\m_{\a,b}$ are a factor of 5-10 below the string scale 
\cite{akr}. Thus they are expected to be in the TeV range.

\section{Calculation of lepton anomalous magnetic moment in the 
         presence of an anomalous $U(1)$}

In order to describe the calculation and its subtleties we will first consider a toy 
model with a photon A, an anomalous $U(1)$ gauge field $B$, chiral charged fermions 
and a complex Higgs. 
We also have an axion $b$ to cancel the anomalies.
We denote by ${\cal L}_{QED}$ the electrodynamics Lagrangian of the photon. 
The relevant part of the low-energy effective Lagrangian can be written as:
\bea
{\cal L} = {\cal L}_{QED} & - &{1\over 4}F^2 + M_s^2(\partial b+ B)^2
                            + D_{\m} H D^{\m} H^* + V(|H|^2)                 \nonumber\\
                          & + & Q_{L} \bar{\psi}_{L} \sla{B} \psi_{L}
                            +   Q_{R} \bar{\psi}_{R} \sla{B} \psi_{R} 
                            + h \bar{\psi}_{L} \psi_{R} H+c.c.               \label{LUone}
\eea
were $B_{\m}$ is the anomalous $U(1)$ with field strength $F_{\m\n}$.
This Lagrangian (\ref{LUone}) is invariant under the ``anomalous" $U(1)$ transformations.
\bea
B^{\m}     & \rightarrow & B^{\m} + \partial^\m \e       \nonumber\\
\psi_L     & \rightarrow & e^{iQ_{L}\e} \psi_L           \nonumber\\
\psi_R     & \rightarrow & e^{iQ_{R}\e} \psi_R           \label{u1trans}\\
H          & \rightarrow & e^{i(Q_{R}-Q_{L})\e}H         \nonumber\\
b          & \rightarrow & b- \e                         \nonumber
\eea
There are two sources of gauge symmetry breaking. One is the stringy mass term and
the other is the non-zero expectation value of the Higgs.
Writing $H=r e^{i \f}$, the Higgs potential fixes the vacuum expectation value $<r>=v$.
The kinetic term of the Higgs field gives an extra contribution to the B mass term:
\be
v^2(\partial \f +\D Q B)^2
\ee

To proceed with the one-loop calculation, it is necessary to add a gauge fixing term 
\be
{\cal L}_{gauge fixing} = \l \Big(\partial B + {c M_s^2 \a\over {\l}} 
- {\Delta Q v^2 \f \over \l}\Big)^2
\label{gaugefixing}
\ee
which keeps $B_\m$ orthogonal to $b$ and $\f$.
Redefining $\tilde{b}=M b$ and $\tilde{\f}=v\f$ we can diagonalize the 
axions doing a $SO(2)$ rotation
\be
\left(\ba {c}
 b'\\
\f'\ea\right)
=
\left(\ba {cc}
 \cos \q' & \sin \q' \\
-\sin \q' & \cos \q' 
\ea\right)
\left(\ba {c}
\tilde{b}\\
\tilde{\f}\ea
\right)
\label{rotSU2}
\ee
where $\cos\q'={cM_s \over \sqrt{c^2 M_s^2 + v^2 \Delta Q^2} }$ and 
$\sin\q'={\Delta Q v \over \sqrt{c^2 M_s^2 + v^2 \Delta Q^2}}$.
Now, the effective Lagrangian has the form
\bea
{\cal L} = {\cal L}_{ED} 
       & - &{1\over 4}F_B^2 + (c^2 M_s^2 + \Delta Q^2 v^2) B^2               \nonumber\\
       & + &(\partial b')^2 + {c^2 M_s^2 + \Delta Q^2 v^2 \over \l} b'^2 
         +  (\partial \f')^2                                                 \nonumber\\
       & + & Q_L \bar{\psi}_L \sla{B} \psi_L
         + Q_R \bar{\psi}_R \sla{B} \psi_R                                   \label{LUfinal}\\
       & + & hv \bar{\psi}_L \psi_R e^{i (\sin\q' b'+\cos\q' \f')/v} + c.c.  \nonumber
\eea
The masses are:
\bea
m_{\psi} &=& hv                                                    \nonumber\\
m_B &=& \sqrt{c^2 M_s^2 + v^2 {\Delta Q}^2}                        \nonumber\\
m_{\f'} &=& 0                                                      \label{masses}\\
m_{b'}  &=& \sqrt{c^2 M_s^2 + v^2 {\Delta Q}^2}/ \sqrt{\l}.        \nonumber
\eea 
We define $m_B=\m$ for simplicity. The Yukawa interactions are given by: 
\vspace{1cm}
\bea
iQ_L \g_\m \Big( {1-\g_5 \over 2} \Big) &,& \quad
\unitlength=0.6mm
\begin{fmffile}{fmfvert1}
\begin{fmfgraph*}(40,25)
\fmfpen{thick}
\fmfleft{i1,i2}
\fmflabel{$\y_L$}{i2}
\fmflabel{$\y_L$}{i1}
\fmfright{o2}
\fmflabel{$B^\m$}{o2}
\fmf{fermion}{i1,v2,i2}
\fmf{photon}{o2,v2}
\end{fmfgraph*}
\end{fmffile}
\nonumber\\[0.7cm]
iQ_R \g_\m \Big( {1+\g_5 \over 2} \Big) &,& \quad
\unitlength=0.6mm
\begin{fmffile}{fmfvert2}
\begin{fmfgraph*}(40,25)
\fmfpen{thick}
\fmfleft{i1,i2}
\fmflabel{$\y_R$}{i2}
\fmflabel{$\y_R$}{i1}
\fmfright{o2}
\fmflabel{$B^\m$}{o2}
\fmf{fermion}{i1,v2,i2}
\fmf{photon}{o2,v2}
\end{fmfgraph*}
\end{fmffile}
\nonumber\\[0.7cm]
{m \D Q \over \m} \g_5 &,& \quad
\unitlength=0.6mm
\begin{fmffile}{fmfvert3}
\begin{fmfgraph*}(40,25)
\fmfpen{thick}
\fmfleft{i1,i2}
\fmflabel{$\y$}{i2}
\fmflabel{$\y$}{i1}
\fmfright{o2}
\fmflabel{$b'$}{o2}
\fmf{fermion}{i1,v2,i2}
\fmf{scalar}{v2,o2}
\end{fmfgraph*}
\end{fmffile}
\nonumber\\[0.7cm]
{h c M_s \over \m} \g_5 &,& \quad
\unitlength=0.6mm
\begin{fmffile}{fmfvert4}
\begin{fmfgraph*}(40,25)
\fmfpen{thick}
\fmfleft{i1,i2}
\fmflabel{$\y$}{i2}
\fmflabel{$\y$}{i1}
\fmfright{o2}
\fmflabel{$\f'$}{o2}
\fmf{fermion}{i1,v2,i2}
\fmf{scalar}{v2,o2}
\end{fmfgraph*}
\end{fmffile}
\nonumber\eea
The propagators are:
\be
D_B^{\m \n}(k) = {-ig^{\m \n}\over {k^2-\m^2} }
             + (1-\l^{-1}) {ik^\m k^\n \over (k^2-\m^2)(k^2-\m^2/{\l})}  \nonumber\ee
\be
G_{\f'}(k)      = {i\over k^2}                                           \label{propagators}\ee
\be
G_{b'}(k)       = {i\over k^2-\m^2/{\l}}                                 \nonumber\ee

To derive the AMM of a lepton, we consider the three-point function of two leptons and 
a photon where a gauge boson or the two scalars can be exchanged on the internal line:
\vspace{.8cm}
\begin{center}
\unitlength=1.2mm
\begin{fmffile}{fmfdf4}
\begin{fmfgraph*}(40,25)
\fmfpen{thick}
\fmfleft{i1,i2}
\fmflabel{$p'$}{i2}
\fmflabel{$p $}{i1}
\fmfright{o2}
\fmflabel{$A^\m$}{o2}
\fmf{fermion}{i1,v1}
\fmf{fermion,label=$p-k$,l.side=right}{v1,v2}
\fmf{fermion,label=$p'-k$,l.side=right}{v2,v3}
\fmf{fermion}{v3,i2}
\fmf{photon}{v2,o2}
\fmffreeze
\fmf{photon,label=$k$,l.side=left}{v1,v3}
\end{fmfgraph*}
\end{fmffile}
\end{center}
We sandwich the above diagram between two on-shell spinors, so we can use the
Gordon decomposition and the mass-shell conditions.
Our goal is to write the expression in the form:
\be
\bar{u}(p') \Big\{ \g_\m F_1(q^2)+{i \s_{\m\n} q^\n \over 2m}F_2(q^2) \Big\} u(p)
\label{goal}
\ee
where $q_\m=p'_\m-p_\m$. The $F_2(q^2=0)$ will give us a correction of the AMM of
the lepton which propagates.
In the present calculation, we have to include diagrams which are coming from
the non trivial couplings between the anomalous $U(1)$s and leptons.
The external vector gauge abelian field is the photon, the internal
propagating fields with momentum $k$ can be the anomalous $U(1)$ gauge boson or the scalars (axions).
We will outline here these calculations. More details can be found in appendix B. 
 
As the anomalous $U(1)$ 
couples differently to left and right leptons, it is neccesary to consider  diagrams
where chirality is conserved (L-L, R-R diagrams) and others where chirality is different
(L-R, R-L). The corresponding diagrams is 
\vspace{.8cm}
\begin{center}
\unitlength=1.2mm
\begin{fmffile}{fmfdf5}
\begin{fmfgraph*}(40,25)
\fmfpen{thick}
\fmfleft{i1,i2}
\fmflabel{$\y_s$}{i2}
\fmflabel{$\y_l$}{i1}
\fmfright{o2}
\fmflabel{$A^\m$}{o2}
\fmf{fermion,label=$p$,l.side=right}{i1,v1}
\fmf{fermion,label=$p-k$,l.side=right}{v1,v2}
\fmf{fermion,label=$p'-k$,l.side=right}{v2,v3}
\fmf{fermion,label=$p'$,l.side=right}{v3,i2}
\fmf{photon}{v2,o2}
\fmffreeze
\fmf{photon,label=$B_{anomalous}(k)$,l.side=left}{v1,v3}
\end{fmfgraph*}
\end{fmffile}
\end{center}
and in algebraic form:
\be
\bar{u}(p')[\int {d^4k\over (2 \pi)^4} (iQ_s \g_\n P_s) {i\over \sla{p'}-\sla{k}-m} \g_\m
                {i\over \sla{p}-\sla{k}-m} (iQ_l \g_\rho P_l) D^{\n\rho}(k)] u(p)
\label{oneloop}
\ee
where $s, l = L, R$ label the chirality.

The propagator of $U(1)$ contains the arbitrary gauge fixing parameter $\l$. In a non-chiral theory 
$\l$ disappears because of the mass-sell conditions of the two spinors which sandwich
the diagrams (\ref{oneloop}).
In a chiral theory, we need the contribution of $b'$ with mass (\ref{masses})
in order to obtain a  gauge invariant result.
We also have to add the one-loop diagrams of $\phi'$. These diagrams are:
\vspace{.8cm}
\begin{center}
\unitlength=1.2mm
\begin{fmffile}{fmfdf9}
\begin{fmfgraph*}(40,25)
\fmfpen{thick}
\fmfleft{i1,i2}
\fmflabel{$\y$}{i2}
\fmflabel{$\y$}{i1}
\fmfright{o2}
\fmflabel{$A^\m$}{o2}
\fmf{fermion,label=$p$,l.side=right}{i1,v1}
\fmf{fermion,label=$p-k$,l.side=right}{v1,v2}
\fmf{fermion,label=$p'-k$,l.side=right}{v2,v3}
\fmf{fermion,label=$p'$,l.side=right}{v3,i2}
\fmf{photon}{v2,o2}
\fmffreeze
\fmf{scalar,label=$axion(k)$,l.side=left}{v1,v3}
\end{fmfgraph*}
\end{fmffile}
\end{center}
where ``axion" stands for $b'$ or $\f'$. In algebraic form they are given by:
\be
{m^2 \Delta Q^2 \over \m^2} \bar{u}(p') \int {d^4k\over (2 \pi)^4} \g_5 {i\over \sla{p'}
             -\sla{k}-m} \g_\m {i\over \sla{p}-\sla{k}-m} \g_5 G_{b'}(k) u(p)
\label{AxionBloop}
\ee
for the $b'$ axion and
\be
{(h c)^2 M_s^2 \over \m^2} \bar{u}(p') \int {d^4k\over (2 \pi)^4} \g_5 {i\over \sla{p'}
             -\sla{k}-m} \g_\m {i\over \sla{p}-\sla{k}-m} \g_5 G_{\phi '}(k) u(p)
\label{AxionFloop}
\ee
for $\f'$.
We expect the sum of the three diagrams to be  $\l$-independent. 
In Appendix B we show this explicitly.
In view of this, we can use any gauge for the evaluation.
For simplicity, 
we choose the Feynman - t'Hooft gauge $\l=1$

The steps of this calculation are as follow:

\noindent
a) Express the denominator as a perfect square using the Feynman parameter trick and 
shifting the loop momentum.

\noindent
b) Move all the $\sla{p'}$ to the left, all the $\sla{p}$ to the right and
make use of the on-shell spinor conditions.

\noindent
c) Perform the momentum integral of the loop after  a Wick rotation to Euclidean space.

\noindent
d) Distinguish terms proportional to $p_\m$ and $p'_\m$.

\noindent
e) Integrate the remaining variables that resulted from Feynman parameter trick.

Following the steps above, we find for the anomalous $U(1)$ exchanged diagram
(details can be found in Appendix B):
For L-L and R-R diagrams:
\be
-{Q_L^2+Q_R^2 \over 16m\pi^2}(p_\m+p'_\m)\int_0^1 dx {x(x^2-3x+2)\over x^2+(1-x){\m^2\over m^2}}
\label{ammchiral}
\ee
For mixed diagrams (L-R and R-L):
\be
-{Q_L Q_R\over 16m\pi^2} (p_\m+p'_\m)\int_0^1 dx {2x(1-x)\over x^2+(1-x){\m^2\over m^2}}
\label{ammmixed}
\ee
The axion $b'$ exchange diagram gives
\be
{\Delta Q^2\over 16m\pi^2} {m^2\over \m^2} (p_\m+p'_\m)\int_0^1 dx 
{x^3\over x^2+(1-x){\m^2\over m^2}}
\label{ammAxionB}
\ee
The diagram for the axion $\f'$ has the same integral with 
(\ref{ammAxionB}) in the limit $\m\rightarrow 0$. Since however the axion is expected to get 
a small mass from non-perturbative effects  
we will consider it with $m_{\f'}$ small. In this case we obtain
\be
{(h c)^2\over 16m\pi^2} {M_s^2\over \m^2} (p_\m+p'_\m)
\int_0^1 dx {x^3\over x^2+(1-x){m_{\f'}^2\over m^2}}
\label{ammAxionFmassive}
\ee
As $M_s/\m\sim 1$, the limit of (\ref{ammAxionFmassive}) for $m_{\f'}\rightarrow 0$ is:
\be
{h^2\over 16m\pi^2} (p_\m+p'_\m){1 \over 2}
\label{ammAxionF}
\ee
\FIGURE[hbt]{
    \includegraphics[height=8cm]{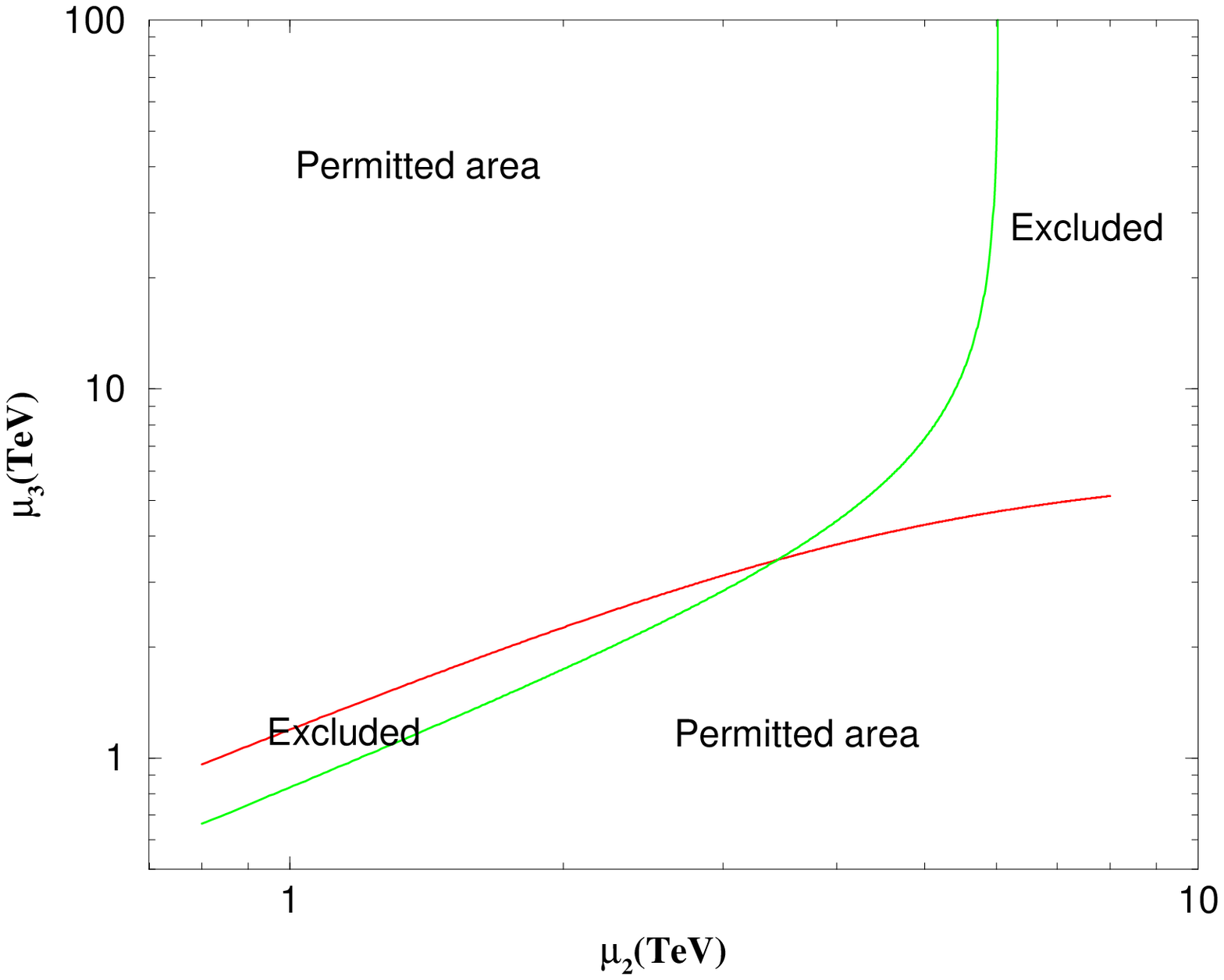}
    \caption{The $z=0$ model. Between the two plots is the excluded area, 
             where the determinant of the second order equation is negative.}
\label{restrictarea1}}
\FIGURE[hbt]{
    \includegraphics[height=8cm]{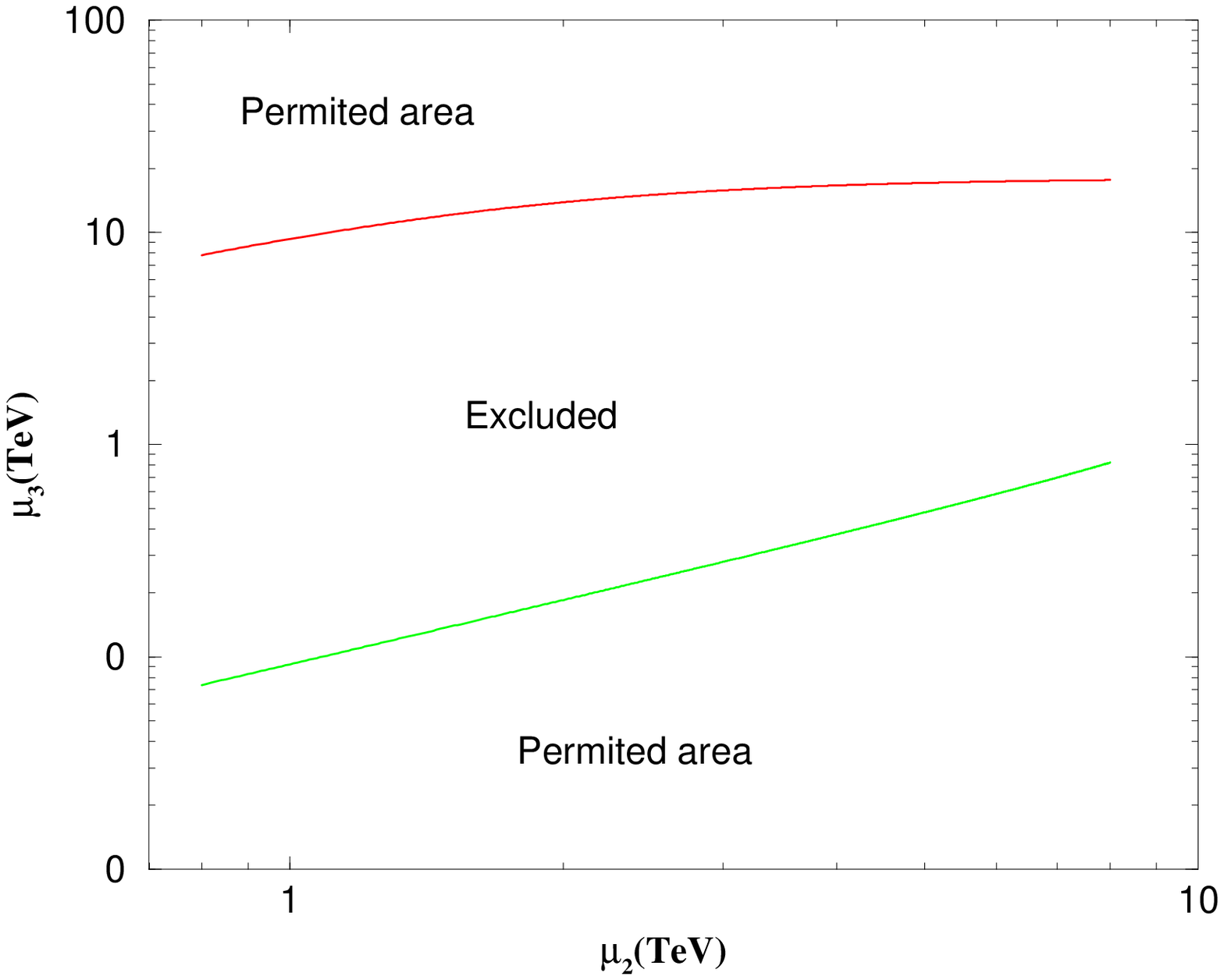}
    \caption{The $z=-1$ model. Between the two plots is the 
             excluded area where the determinant of the second order 
             equation is negative.}
\label{restrictarea2}}

\section{Anomalous magnetic moment of muon in the D-brane 
         realization of the standard model}

Using the results above we can now embark in the calculation of the AMM of the muon
in the D-brane realization of the SM. To do this we have to include the 
contribution of (\ref{ammchiral}) and (\ref{ammmixed}) for both anomalous $U(1)$s 
as well as the (\ref{ammAxionB}) and (\ref{ammAxionF}) for the axion diagrams to the
SM result\footnote{We use for simplicity $m=m_{muon}$.}.
\bea
\d\a= & {1\over 8\p^2}\sum_{i=\a,b}\Big({m\over \m_i}\Big)^2
\int_0^1 dx {x(m^2\Delta Q_i^2 x^2+\m_i^2(4 Q_{iL} Q_{iR}-(2-x)(Q^2_{iL}+ Q^2_{iR}))(1-x)
\over m^2 x^2+\m_i^2(1-x^2)}                 \nonumber\\
 & +{h^2\over 16\p^2} 
\label{FullAMM}
\eea
In our case $\m_i \gg m$, therefore we expand the contributions and 
keep the terms up to second order in $(\m_i / m)$.
The final result is
\be
\a^{U(3)\times U(2) \times U(1)}_{muon}=\a^{SM}_{muon}
+\sum_{i=\a,b} {Q_{i L}^2-3Q_{i L}Q_{i R}+Q_{i R}^2\over 12\p^2} 
\Big({m\over \m_i}\Big)^2 +{h^2\over 16\p^2} 
\label{AMM}
\ee
where $Q_{\a L}, Q_{\a R}, Q_{b L}, Q_{b L}$ are the rotated by (\ref{rotchargone}) 
or (\ref{rotchargtwo}), charges of (\ref{charges}).
We use as $Q_{iL}$ and $Q_{iR}$ the charges of the $L$ and $l^c$ in (\ref{charges}).

Using the measured difference (\ref{difference}) we can express one of the unknown variables as a 
function of the two others. Thus, for $z=0$ we can find the $\m_\a$ and $\m_\b$ dependence 
of $\tan \q$. We have to solve a second order equation:
\bea
(12\p^2 \m_\a^2 \m_b^2 (\d\a-\a_{\f'})+m^2(817\m_\a^2 -1220\m_b^2))\tan^2\q
+26\sqrt{215}m^2(\m_\a^2 -\m_b^2)\tan\q &           \nonumber\\
+12\p^2 \m_\a^2 \m_b^2 (\d\a-\a_{\f'}) -1220 m^2 \m_\a^2 +817 m^2\m_b^2&=0
\nonumber\\  
& \label{tanthetaFir}
\eea
where we denote as $\a_{\f'}$ the contribution from the axion $\f'$. As $\tan \q$ is real, 
the discriminant must be positive. We can easily find the excluded area in the $\m_2$,$\m_3$ 
plane where this discriminant is negative. In Fig. \ref{restrictarea1} we plot this area 
for the z=0 model.

For the $z=-1$ model we obtain
\bea
(12\p^2 \m_\a^2 \m_b^2 (\d\a-\a_{\f'}) -m^2(10363\m_\a^2 +580\m_b^2))\tan^2\q
-362\sqrt{215}m^2(\m_\a^2 -\m_b^2)\tan\q &           \nonumber\\
+12\p^2 \m_\a^2 \m_b^2 (\d\a-\a_{\f'}) -m^2(580\m_\a^2 +10363m^2\m_b^2)&=0
\nonumber\\  
& \label{tanthetaSec}
\eea
and the allowed area is plotted in Fig.\ref{restrictarea2}.
As mentioned before the anomalous $U(1)$ masses are  expected to be in the TeV range.
Thus, there is little allowed space in this case in order to reproduce the experimental result.

Until now we have evaluated diagrams of the lowest lying  string states. 
The massive oscillator 
string states at level n have masses equal to $\sqrt{n}M_s$. 
The ratio of the contribution of such a state to 
that of a low lying state is expected to scale as the square of the ratio 
of the masses. Thus corrections due to the first massive level are in the 
1-5\% range and higher levels are 
further suppressed.
There are also KK states that can contribute. However their masses as mentioned
earlier are as large as the string scale and thus give suppressed contributions.
A direct string calculation is under way in order to corroborate 
the expectations above.

\section{Conclusion}

In this paper we have analysed contributions to the anomalous magnetic moment of leptons 
in the minimal D-brane realization of the Standard Model.
We have shown that the two anomalous massive gauge bosons present 
\cite{akt} with masses 
in the TeV range, provide contributions that have the correct order
 of magnitude to accommodate 
the recent experimental data \cite{BNL}.
Further contributions from string oscillators and KK states 
are expected to be 
sufficiently suppressed. 
A string calculation is however necessary in order to verify this expectation.

It is an important open problem to find an orientifold vaccum of string theory
that realizes the standard model as described in \cite{akt} with the correct
tree level couplings.
In such a case all relevant observable quantities could be calculated
precisely.

\vskip 1.5cm
\centerline{\bf\Large Acknowledgments}
\vskip .5cm

The authors would like to thank I. Semertzidis for a communication 
and I. Antoniadis, C. Coriano and A. Pilaftsis for comments
on the manuscript. They would also like to thank the Laboratoire de Physique Th{\'e}orique de 
l'Ecole Polytechnique and the Laboratoire de Physique Th{\'e}orique de
l'Ecole Normale Superieure for hospitality during the last stages of this work.
This work was partially supported by
RTN contracts HPRN--CT--2000-00122 and --00131.

\bigskip\appendix\section{The extended Standard Model fields}

In this appendix we provide some more details about the masses of the fields and the gauge 
couplings. Based on (\ref{Higgs}) the Higgs expectation values  have the form: 
\bea
h={v\over \sqrt{2}}
  \left(\ba {c} 1\\ 0 \ea\right) & , &
\bar{h}={v\over \sqrt{2}}
        \left(\ba {c} 1\\ 0 \ea\right) .
\label{vev}
\eea
Thus, the covariant derivative of the Higgs (in the $z=0$ model) is
\be
D^{\m} H={v\over \sqrt{2}}(\partial^\m-i{g_3 \bf{1} \over \sqrt{2}}A_1^\m
          -i{g_2 \bf{1} \over 2}A_2^\m -i{g_2 \over 2} \t_\a W^\m_\a )
          \left(\ba {c} 1\\ 0 \ea\right) e^{i \f}
\ee
\be
D^{\m} H'={v\over \sqrt{2}}(\partial^\m +i{g_2 \bf{1}\over 2}A_2^\m 
           -i{g_2 \over 2} \t_\a W_\a^2 )
           \left(\ba {c} 1\\ 0 \ea\right) e^{i \f'}
\ee
where $W_\a$, $\a=1,2,3$ the $SU(2)$ gauge bosons. We normalize all $U(N)$
generators according to $Tr T^\a T^b=\d^{\a b}/2$ and measure the corresponding
$U(1)_N$ charges with respect to the coupling $g_N/\sqrt{2N}$, with $g_N$ the 
$SU(N)$ coupling constant as in \cite{akt}. We have also $g_1=g_3$.

The mass matrix for the gauge bosons is
\be
M=V^T m V
\ee
where $V^T=(A_1,A_2,A_3,W_3,W_1,W_2)$ and 
\be
m={v^2\over 4} \left(\ba {cccccc} 
      g_3^2 & {g_2 g_3 \over \sqrt{2}} & 0 &  {g_2 g_3 \over \sqrt{2}} & 0     & 0 \\
    {g_2 g_3 \over \sqrt{2}} & g_2^2          & 0 & 0                  & 0     & 0 \\ 
                           0         &    0   & 0 & 0                  & 0     & 0 \\
            {g_2 g_3 \over \sqrt{2}} &    0   & 0 & g_2^2              & 0     & 0 \\
                    0                &    0   & 0 & 0                  & g_2^2 & 0 \\
                    0                &    0   & 0 & 0                  & 0     & g_2^2

\ea\right).
\label{bosonmassmatrix}
\ee
Doing a rotation with the matrix (\ref{Uarrey}), we can go to a basis where 
$\tilde{A_1}$ is the hypercharge. The other two $U(1)$ bosons 
$\tilde{A_2}, \tilde{A_3}$ are anomalous and we expect two axions $\a_2, \a_3$
to cancel the anomalies. Inserting
\be
{\cal L}_{axionic terms} =
{1\over 2}(\partial \a_2 -M_2 \tilde{A_2})^2 
+{1\over 2}(\partial \a_3 -M_3 \tilde{A_3})^2,
\ee
two elements of the rotated mass matrix will be shifted. Since $v\ll M_2,M_3\sim M_s$, we can 
perturbatively diagonalize this matrix and find the new masses of these new 
fields. Finally, there is a massless state (photon), a ``light" $Z$ boson with mass
\be
m^2_Z={v^2 g_2^2 r^2 \over 2 t^2}
-v^4{g_2^2 g_3^2 r^2 s^2 (M_2^2+M_3^2+(M_2^2-M_3^2)cos 2\q)
\over 64 t^4 M_2^2 M_3^2} +O \Big[ {M_Z^6 \over M_s^4}\Big]
\ee
and two heavy ones with masses:
\bea
\m^2_2     =& M_2^2+v^2{8g_2^4 t^2 cos^2\q 
                   +g_3 sin\q (-4g_2^2 t^3 cos\q 
                   +g_3 (130g_2^4+66 g_2^2 g_3^2 +9 g_3^4)sin\q)
                        \over 2 s^2 t^2} +O \Big[ {M_Z^4 \over M_s^2}\Big] \nonumber\\
\m^2_3     =& M_3^2+v^2{g_3^2(130g_2^4+66 g_2^2 g_3^2 +9 g_3^4)cos^2\q 
                   +4 g_2^2 g_3 t^3 cos\q sin\q +8 g_2^4 t^2 sin^2\q
                        \over 2 s^2 t^2} +O \Big[ {M_Z^4 \over M_s^2}\Big]
\eea
where $t=\sqrt{14 g_2^2+3 g_3^2}$, $s=\sqrt{16 g_2^2+6 g_3^2}$,
$f=\sqrt{11 g_2^2+3 g_3^2}$, $r=\sqrt{7g_2^2 +3g_3^2}$ and $\m_i=m_{A'_i}$, the
masses of the new anomalous U(1)s.
The old fields as functions of the new rotated fields are:
\bea
A_1 \approx&  {2 \sqrt{3} t g_2 A'_1 
              - \sqrt{2} r s sin\q A'_2
              + \sqrt{2} r s cos\q A'_3 
              - 6 g_2 g_3  W'_3  \over 2 t r}                    \nonumber\\
A_2 \approx&  {-\sqrt{6}g_3 s t A'_1 
              + 4 g_2 r (2 t cos\q -3 g_3 sin\q) A'_2
              + 4 g_2 r (3 g_3 cos\q +2 t sin\q) A'_3 
              + 3\sqrt{2} g_3^2 s W'_3 \over 2 r s t}            \nonumber\\
A_3 \approx&  {2 g_2 s t A'_1 
              + \sqrt{6} r (g_3 t cos\q +4 g_2^2sin\q) A'_2
              + \sqrt{6} r (-4g_2^2 cos\q +g_3 t sin\q) A'_3 
              - 2\sqrt{3} g_2 g_3 s W'_3  \over r s t}           \label{relations}\\
W_3 \approx& -{\sqrt{3} g_3 A'_1 + t W'_3 \over \sqrt{2} r}      \nonumber
\eea
where $A'_1$ and $W'_3$ are the photon and the $Z^0$.

It is necessary to add a $R_{\x}$ gauge fixing term. This will cancel some
mixing terms which are coming from the kinetic terms of the Higgses and it will 
maintain the manifest unitarity of the theory with spontaneously broken gauge symmetry.
\bea
{\cal L}_{gauge fixing} &=& \l (\partial A'_1)^2                       \nonumber\\
                        & & +\m \Big(\partial A'_2- 
                            v^2{ 2(\f-\f')g_2^2 t cos\q -g_3(f^2 \f -3g^2_2 \f')sin\q
                            \over 2 \m t s}
                            -{M_2\over 2\m}\a_2 \Big)^2                \nonumber\\
                        & & +\r \Big(\partial A'_3- 
                            v^2{g_3(f^2 \f -3\f' g_2^2)cos\q +2(\f-\f')g_2^2 t sin\q
                            \over 2 \r t s}
                            -{M_3\over 2\r}\a_3 \Big)^2                \nonumber\\
                        & & +\s \Big(\partial W'_3+
                            v^2{(\f+\f')g_2 r \over 2 \sqrt{2} \s t}\Big)^2
\eea
The gauge fixing terms give masses to the axions and to the Higgs. We
can diagonalize perturbatively the mass-matrix of these fields. Considering
$\m=\l=\r=\s$ we find one massless and three massive fields:
\bea
m^2_{\tilde{a}_2} & = & {M_2^2 \over 4 \m}+O[M_Z^2]  \nonumber\\
m^2_{\tilde{a}_3} & = & {M_3^2 \over 4 \m}+O[M_s^2]  \label{NHAmasses}\\
m^2_{\tilde{\f}}  & = & {1 \over 4 \m}{g_2^2 r^2 v^4 \over t^2}
                   +O \Big[ {M_z^2 \over M_s^2}\Big] \nonumber\\
m^2_{\tilde{\F}}  & = & 0                            \nonumber
\eea
The old fields as a functions of the new ones are:
\bea
\a_2 \approx & \tilde{a}_2
               -{v^4 (4 g_2^2 g_3 t^3 cos2\q 
               +(112 g_2^6-106 g_2^4 g_3^2-66 g_2^2 g_3^4-9 g_3^6)sin2\q)
               \over 2 t^2 s^2 M_2 M_3} \tilde{a}_3
               +{v^2 g_3 s^2 sin\q \over \sqrt{2} t M_2} \tilde{\f}
               +{v^2 (4 g_2^2 cos\q - g_3 t sin\q \over \sqrt{2} s M_2} \tilde{\F} 
               \nonumber\\
\a_3 \approx & \tilde{a}_3
               -{v^2 g_3 s^2 cos\q \over \sqrt{2} 2 t M_3} \tilde{\f}
               +{v^2 (g_3 t cos\q + 4 g_2^2 sin\q \over \sqrt{2} s^2 M_3} \tilde{\F} 
               \nonumber\\
\f   \approx & {v^2(2 g_2^2 t cos\q -g_3 f^2 sin\q)\over t s M_2}\tilde{a}_2
               +{v^2(g_3 f^2 cos\q +2 g_2^2 t sin\q)\over t s M_3}\tilde{a}_3
               +{1 \over \sqrt{2}}\tilde{\f}
               -{1 \over \sqrt{2}}\tilde{\F}
               \nonumber\\
\f'  \approx & {v^2 g_2^2 (-2 t cos\q +3 g_3 sin\q)\over t s M_2}\tilde{a}_2
               -{v^2 g_2^2 (3 g_3 cos\q +2 t sin\q)\over t s M_3}\tilde{a}_3
               +{1 \over \sqrt{2}}\tilde{\f}
               +{1 \over \sqrt{2}}\tilde{\F}
\eea
From the trilinear Yukawa couplings we can find how leptons couple to
the new Higgses and axions. Using (\ref{HY}), we find the following vertices:
\vspace{1cm}
\bea
{h v^3 (2 g_2^2 t cos\q -g_3 f^2 sin\q )\over t s M_2}\g_5 &,& \quad
\unitlength=0.6mm
\begin{fmffile}{fmfvertA1}
\begin{fmfgraph*}(40,25)
\fmfpen{thick}
\fmfleft{i1,i2}
\fmflabel{$\y$}{i2}
\fmflabel{$\y$}{i1}
\fmfright{o2}
\fmflabel{$\tilde{a}_2$}{o2}
\fmf{fermion}{i1,v2,i2}
\fmf{scalar}{v2,o2}
\end{fmfgraph*}
\end{fmffile}
\nonumber\\[0.7cm]
{h v^3 (g_3 f^2 cos\q + 2 g_2^2 t sin\q)\over t s M_3}\g_5 &,& \quad
\unitlength=0.6mm
\begin{fmffile}{fmfvertA2}
\begin{fmfgraph*}(40,25)
\fmfpen{thick}
\fmfleft{i1,i2}
\fmflabel{$\y$}{i2}
\fmflabel{$\y$}{i1}
\fmfright{o2}
\fmflabel{$\tilde{a}_3$}{o2}
\fmf{fermion}{i1,v2,i2}
\fmf{scalar}{v2,o2}
\end{fmfgraph*}
\end{fmffile}
\nonumber\\[0.7cm]
{h v \over \sqrt{2}} \g_5 &,& \quad
\unitlength=0.6mm
\begin{fmffile}{fmfvertA3}
\begin{fmfgraph*}(40,25)
\fmfpen{thick}
\fmfleft{i1,i2}
\fmflabel{$\y$}{i2}
\fmflabel{$\y$}{i1}
\fmfright{o2}
\fmflabel{$\tilde{\f}$}{o2}
\fmf{fermion}{i1,v2,i2}
\fmf{scalar}{v2,o2}
\end{fmfgraph*}
\end{fmffile}
\nonumber\\[0.7cm]
-{h v \over \sqrt{2}} \g_5 &,& \quad
\unitlength=0.6mm
\begin{fmffile}{fmfvertA4}
\begin{fmfgraph*}(40,25)
\fmfpen{thick}
\fmfleft{i1,i2}
\fmflabel{$\y$}{i2}
\fmflabel{$\y$}{i1}
\fmfright{o2}
\fmflabel{$\tilde{\F}$}{o2}
\fmf{fermion}{i1,v2,i2}
\fmf{scalar}{v2,o2}
\end{fmfgraph*}
\end{fmffile}
\nonumber\eea
where $h$ the Yukawa coupling  of $H^\dagger Ll^c$.

\section{The evaluation of lepton vertex functions}

Here we will give some details about the calculation of the lepton AMM.
Our goal is to separate from the vertex functions, terms proportional to $\s^{\m\n} q_m$.
As the vertex functions are sandwiched by two on-shell spinors we can use the Gordon
decomposition and try to distinguish terms proportional to $p^\m$ and $p'^\m$.
We will begin with (\ref{oneloop}) for the anomalous $U(1)$ diagram. We rewrite
it here:
\be
\bar{u}(p')[\int {d^4k\over (2 \pi)^4} (iQ_s \g_\n P_s) {i\over \sla{p'}-\sla{k}-m} \g_\m
                {i\over \sla{p}-\sla{k}-m} (iQ_l \g_\rho P_l) D^{\n\rho}(k)] u(p)
\label{oneloopApp}
\ee
where $s, l = L, R$ denote the chiralities. The propagator of the $U(1)$ $D^{\m\n}$ contains the 
gauge fixing parameter $\l$. This parameter is expected to disappear from physical gauge invariant 
couplings.
We will verify explicitly here that $\l$ disappears from the sum of all the vertex functions.
The $D^{\m\n}$ consist of two terms, one independent and one dependent on $\l$. 
First, we will calculate the correction from the $\l$-independent part. In this case
we have a fraction with three factors in the denominator. 
Using the Feynman parameter trick we write the denominator as follows:
\be
{1\over ((p'-k)^2-m^2)((p-k)^2-m^2)((k^2-\m^2)} = 
          2! \int_0^1 dx \int_0^{1-x} dy {1\over D^3}
\label{FeynmanTrickApp}
\ee
where
\be
D = k^2-2k(px+p'y)+p^2x+p'^2y-m^2(x+y)-\m^2(1-x-y)
\ee
In order to express the denominator as a function of the norm of the momentum, 
we shift $k$ to $k+px+p'y$. We find $D = k^2-\Delta$ where
\be
\Delta = m^2(x+y)+\m^2(1-x-y)
\label{denominatorApp}
\ee

Next, we will express the numerator of (\ref{oneloopApp}) in terms of $k^\m$ in 
order to integrate on the internal momenta. Because of the symmetry, two identities 
are useful here:
\be
\int {d^4k\over (2 \pi)^4} {k^\m \over D^3} = 0
\label{idoddApp}
\ee
\be
\int {d^4k\over (2 \pi)^4} {k^\m k^\n \over D^3} = 
\int {d^4k\over (2 \pi)^4} {{1\over 4}k^2 g^{\m\n} \over D^3}
\label{idevenApp}
\ee
We keep only terms proportional to even powers of $k^\m$. We will separate 
chiral and mixed diagrams:

\noindent
(1) $L-L$, $R-R$ diagrams. The numerator of (\ref{oneloopApp}) with $s=l$ has the form
\be
\g_\n {1\pm\g_5 \over 2} (\sla{A}+m)\g_\m(\sla{C}+m) \g^\n {1\pm\g_5 \over 2}
\label{numeratorStyleApp}
\ee
which, after some algebra becomes
\be
{1\over 2}\g_\n \sla{A}\g_\m \sla{C} \g^\n + {1\over 2} m^2 \g_\n \g_\m \g^\n.
\label{numIApp}
\ee
Terms which contain one $\g_5$ are orthogonal to $\g_{\m\n}$ and we can ignore them. 
Also the second term in (\ref{numIApp}) does not contribute since it is proportional to $\g_\m$. 
Thus, only the first term remains. Shifting $k$ to $k+px+p'y$ we obtain
\be
\g^\n ((1-y)\sla{p'}-x \sla{p}-\sla{k}) \g_\m ((1-x)\sla{p}-y \sla{p'}-\sla{k}) \g_\n
\label{numIIApp}
\ee
Moving all $\sla{p'}$ to the left, all $\sla{p}$ to the right, using (\ref{idoddApp}),
(\ref{idevenApp}) and on-shell conditions, we find
\be
4m[(1-2x-y+xy+x^2)p_{\m} + (1-x-2y+xy+y^2)p_{\m}]
\label{numeratorResalt1App}
\ee
Here there is a symmetry under the reflection $x\leftrightarrow y$. Thus, we can make
the coefficients of $p_{\m}$ and $p'_{\m}$ equal adding the ``reflected" terms and divide
the result by 2. Now, only the integrals on $x$ and $y$ remain. Integrating on $x$
and making a change of variables, we find:
\be
-{Q_s^2\over 16m\pi^2}(p_\m+p'_\m)\int_0^1 dx {x(x^2-3x+2)\over x^2+(1-x){\m^2\over m^2}}
\label{ammchiralApp}
\ee
Our main interest is for $\m\gg m$. Expanding, we find:
\be
{Q_s^2\over 16m\pi^2}(p_\m+p'_\m)
     \Big(-{2\over 3}\Big({m\over \m}\Big)^2
         +\Big(-{19\over 12} -2\log\Big({m\over \m}\Big)\Big)\Big({m\over \m}\Big)^4 
         +O\Big({m\over \m}\Big)^5\Big)
\label{ammchiralexpandApp}
\ee

\noindent
(2) $L-R$ and $R-L$ diagrams. The only difference from the above lies in the numerator.
Working similarly, for $s\neq l$ in (\ref{oneloopApp}) we find
\be
4m[(1-2x)p_{\m} + (1-2y)p'_{\m}]
\label{numeratorResalt2App}
\ee
and finally
\be
-{Q_L Q_R\over 16m\pi^2} (p_\m+p'_\m)\int_0^1 dx {2x(1-x)\over x^2+(1-x){\m^2\over m^2}}
\label{ammmixedApp}
\ee
The expansion for $\m\gg m$ gives: 
\be
{Q_L Q_R\over 16m\pi^2}(p_\m+p'_\m)
     \Big(2\Big({m\over \m}\Big)^2
         -2\Big(-{11\over 3} -4\log\Big({m\over \m}\Big)\Big)\Big({m\over \m}\Big)^4 
         +O\Big({m\over \m}\Big)^5\Big)
\label{ammmixedexpandApp}\ee

We will now calculate the contribution of the second ($\l$-dependent) term of the 
massive gauge field's propagator (\ref{propagators}) in (\ref{oneloop}).
The denominator contains four factors. We will use again the Feynman parameter
trick.
  
Due to the projection operators, there are terms with two, one and no $\g_5$. 
Terms with one $\g_5$ do not contribute to (\ref{goal}) being orthogonal  to both $\g_\m$, 
$\s_{\m\n}$. 
Terms without $\g_5$ vanish using mass-shell conditions of the fermions that 
sandwich the diagram.
Only terms with two $\g_5$'s remain. After a lot of Diracology we obtain
\bea
-(1&-&\l^{-1}){\Delta Q^2 (p_\m+p'_\m) \over 16 \p^2}
\int_0^1 dx \int_0^x dy \int_0^y dz  \times \nonumber\\
& \Bigg( &
-{m (-1 + 3 z) \over m^2 y^2 + \m^2 \Big( x-y+ {1-x \over \l} \Big)}
+{m^3 z y^2 \over \Big( m^2 y^2 + \m^2 \Big( x-y+ {1-x \over \l} \Big) \Big)^2} \Bigg)
\label{secldeptermApp}
\eea

Now, we will calculate the axion diagrams (\ref{ammAxionB}) and (\ref{ammAxionF}). 
The $\b'$ axion diagram is equal to 
\be
{m^2 \Delta Q^2 \over \m^2} \bar{u}(p') \int {d^4k\over (2 \pi)^2} \g_5 {i\over \sla{p'}
             -\sla{k}-m} \g_\m {i\over \sla{p}-\sla{k}-m} \g_5 G_{b'}(k) u(p)
\label{AxionBloopApp}
\ee
The only difference with the $U(1)$ diagram (\ref{oneloopApp}) is in the numerator. 
So, we focus on it and the result is
\be
2[(x^2+yx)p_{\m} + (y^2+xy)p'_{\m}].
\label{numAxionApp}\ee
Thus, the (\ref{AxionBloopApp}) contribution is
\be
{\Delta Q^2\over 16m\pi^2} {\m^2\over m^2} (p_\m+p'_\m)
\int_0^1 dx {x^3\over x^2+(1-x){\m^2\over \l m^2}}
\label{ammAxionBApp}
\ee
In the entire contribution only (\ref{secldeptermApp}) and (\ref{ammAxionBApp}) are $\l$
dependent. Adding these two terms and calculating the $\l$ derivative  using Mathematica we find zero. 
Thus, $\l$ disappears as it should and we can use the Feynman - t'Hooft gauge for simplicity.
As we are interested in $\m\gg m$, we expand (\ref{ammAxionBApp}):
\be
{\Delta Q^2\over 16m\pi^2} {\m^2\over m^2} (p_\m+p'_\m)
     \Big(\Big(-{11\over 6} -2\log\Big({m\over \m}\Big)\Big)\Big({m\over \m}\Big)^4 
     +O\Big({m\over \m}\Big)^5\Big).
\label{ammAxionexpandBApp}
\ee

Let  us now turn to the $\f'$ diagram. The corresponding integral is the $\m\rightarrow 0$ 
limit of the the integral in (\ref{ammAxionBApp}). However we will consider a more 
general case where $\m$ is small. Keeping the same coupling constant as the above
we have
\be
{h^2\over 16m\pi^2}(p_\m+p'_\m)
\int_0^1 dx {x^3\over x^2+(1-x){m_{\f'}^2\over m^2}}
\label{ammAxionFmassiveApp}
\ee
Considering $m_{\f'}$ very small we can expand (\ref{ammAxionFmassiveApp}) and we
find
\be
{h^2\over 16m\pi^2}(p_\m+p'_\m) \Big({1\over 2} + 
\Big(1+ log\Big({m_{\f'}\over m} \Big) \Big) \Big({m_{\f'}\over m}\Big)^2 \Big)
+O \Big( {m_{\f'}\over m} \Big)^3
\label{ammAxionexpandFApp}
\ee
In the last formula there is $h$ which is computable from SM. From (\ref{masses}) 
is obvious that we need to estimate the expectation value of the Higgs $v$. Using
the mass of $Z^0$ $M_{Z^0}=91.19GeV$, the electron charge $e$ and the value of
$sin^2\q_W=0.23$ from SM we find $v=2 M_{Z^0} sin\q_W \sqrt{1-sin^2\q_W}/ e$ so
\be
h={e m_{muon} \over 2 M_{Z^0} sin\q_W \sqrt{1-sin^2\q_W}}
\label{hcoupling}
\ee
%
%


\end{document}